\documentclass[aps,prl,showpacs,twocolumn,superscriptaddress]{revtex4}
\usepackage{bm,color}
\usepackage{graphicx}
\usepackage{amsmath, amssymb}
\usepackage{braket}
\usepackage{comment}
\begin{document}
\title{Photoinduced 120-degree spin order in the Kondo-lattice model on a triangular lattice}
\author{Takashi Inoue}
\affiliation
{Department of Applied Physics, Waseda University, Okubo, Shinjuku-ku, Tokyo 169-8555, Japan}
\author{Masahito Mochizuki}
\affiliation
{Department of Applied Physics, Waseda University, Okubo, Shinjuku-ku, Tokyo 169-8555, Japan}
\begin{abstract}
We theoretically predict the emergence of 120-degree spin order as a nonequilibrium steady state in the photodriven Kondo-lattice model on a triangular lattice. In the system away from the half filling with ferromagnetic ground state, the photoexcitation of conduction electrons and the photoinduced renormalization of bandwidth cause reconstruction of band structure and subsequent redistribution of the electrons through relaxations, which result in an electronic structure similar to that in the half-filled system at equilibrium, where the 120-degree spin order is stabilized. In this photoinduced 120-degree spin ordered phase, domains of different spin-ordered planes are formed, and vortices of spin chirality vectors called $Z_2$ vortices appear at points where multiple domains meet. We also discuss favorable conditions and experimental feasibility to observe the predicted photoinduced magnetic phase transition by investigating dependencies on the light parameters (amplitude, frequency, and polarization), the electron filling, the strength of Kondo coupling, and the effects of antiferromagnetic coupling among the localized spins. Photoinduced magnetic structures proposed so far have been limited to simple collinear (anti)ferromagnetic orders or local magnetic defects in magnets. The present work paves a way to optical creation of complex noncollinear magnetisms as global nonequilibrium steady phase in photodriven systems.
\end{abstract}

\maketitle
\section{INTRODUCTION}
\begin{figure}[tbh]
\centering
\includegraphics[scale=0.5]{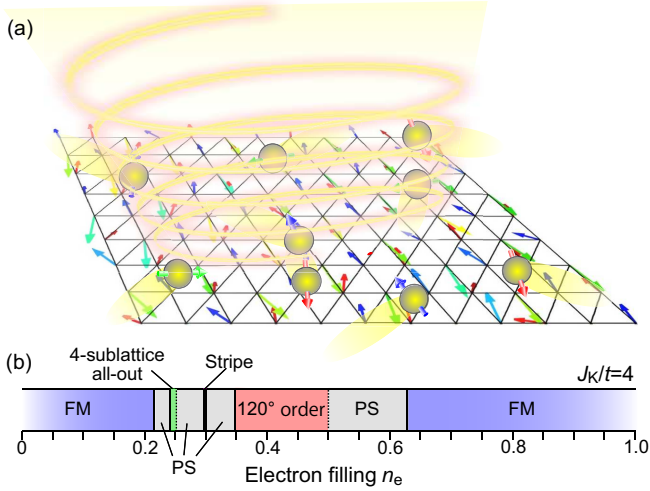}
\caption{(a) Schematics of photoirradiated Kondo-lattice system. (b) Ground-state phase diagram of the Kondo-lattice model in Eq.~(\ref{eq:Eq01}) on a triangular lattice as a function of the electron filling $n_{\rm e}$ reproduced from Ref.~\cite{Akagi2010}. The tight-binding term of the model contains only the nearest-neighbor transfer integrals $t$, while the Kondo-coupling constant $J_{\rm K}$ is set to be $J_{\rm K}/t=4$. Ferromagnetic (FM) phases appear in the lower and higher filling regimes of $n_{\rm e} \leq 0.22$ and $n_{\rm e} \geq 0.63$, while the 120-degree spin order appears in the regime near the half filling, i.e., $0.35 \leq n_{\rm e} \le 0.5$. Phase separation (PS) occurs in areas sandwiched by the FM and 120-degree ordered phases. A four-sublattice all-out phase and a stripe phase also appear inside the PS region.}
\label{Fig01}
\end{figure}
States of matters and their variations triggered by external stimuli often provide us interesting problems of physics and useful device functions. Along with a recent development of the laser technology, the research field on optical control of states in materials is now growing rapidly~\cite{Kirilyuk2010,Aoki2014,Mentink2017,Basov2011,Barman2020,CWang2020}. An advantage of the usage of optical means as compared with other techniques is quick responses and fast time scales. In addition, the optical control enables us to reduce energy consumption by taking advantage of its resonance nature and even to realize wear-free devices because it is a contactless technique. 

The optical manipulation of magnets is one of the most important subjects in this field because the magnetisms have numerous functionalities applicable to electronics devices as intensively studied in the field of spintronics nowadays~\cite{Satoh2010,Mertelj2010,Razdolski2017,Radu2011,Fiebig1998,Averitt2001,Rini2007,Ichikawa2011,Zhao2011,Yada2016,Koshihara1997,Mikhaylovskiy2020,Lin2018,Mentink2015,Chovan2006,Matsueda2007,Koshibae2009,Koshibae2011,Ohara2013,Kanamori2009,PWang2020,Ono2017,Ono2018,Ono2019}. One of the interesting phenomena is the light-induced demagnetization where thermal fluctuations induced by heat of applied laser light melts magnetic orders and renders the system paramagnetic~\cite{Beaurepaire1996}. Another interesting phenomenon is photoinduced ferromagnetism~\cite{Koshibae2009,Kanamori2009} which can be achieved in double-exchange systems~\cite{Zener1951,Anderson1955,Gennes1960} through metallization with the light-induced electron excitations. These photoinduced magnetic states appear as nonequilibrium steady phases in the photodriven systems. However, most of the photoinduced magnetically ordered phases are limited to simple magnetisms such as collinear (anti)ferromagnetic orders, local magnetic defects, and paramagnetic state.

On the other hand, several noncollinear magnetisms such as ferromagnetic domain walls, skyrmions, merons, hedgehogs, and bubbles have turned out to host a lot of interesting physical phenomena and potentially useful device functions, which are now subject to intensive studies. Optical creations of such noncollinear magnetisms have also attracted much attention, and several methods have been proposed both experimentally~\cite{Je2018,Ogawa2015,Gerlinger2021} and theoretically~\cite{Tsesses2018,Stepanov2017,Fujita2017,Yang2018,Yang2019}. However, most of the proposed methods are based on local application of heats or alternate-current (ac) electromagnetic fields, and the noncollinear magnetism appears not as a thermodynamic phase but as individual defects in a uniform ferromagnetic state. To our knowledge, there have been neither theoretical proposals nor experimental demonstrations for optical creation of noncollinear magnetic order as a global nonequilibrium steady phase.

In this paper, we theoretically demonstrate that the 120-degree spin order can be created by irradiating the Kondo-lattice system on a simple triangular lattice by laser light. Starting with a uniform ferromagnetic state away from the half filling, we reveal that the photoirradiation causes excitations of conduction electrons and subsequent band-gap formation through narrowing the bandwidth. After the reconstruction of band structure, the conduction electrons excited to upper-lying states above the gap are relaxed to fall to lower-lying states below the gap, and they are eventually distributed uniformly in the lower-lying states. Through these processes, an electronic structure similar to that in the half-filled system takes place in the photodriven system. In such a system, the conduction electrons are subject to nesting vectors of the Fermi surface(s) similar to those in the half-filled system at equilibrium where the 120-degree spin order is stabilized. This mechanism is expected in a wide variety of spin-charge coupled magnets described by the Kondo-lattice models and the double-exchange models. Thereby, the present work will realize a breakthrough in the research of optical control of magnetism, which has been attracting intensive interest recently.

\section{FORMULATION}
\subsection{A. Model}
We start with a Kondo-lattice model on a triangular lattice. The Hamiltonian at equilibrium is given in the form,
\begin{equation}
\mathcal{H}=-\sum_{\langle ij \rangle\sigma} t_{ij}c^\dag_{i\sigma}c_{j\sigma}
-J_\mathrm{K}\sum_{i\sigma\sigma'}\bm{S}_i \cdot c^\dag_{i\sigma}\bm{\sigma}
_{\sigma\sigma'}c_{i\sigma'}.
\label{eq:Eq01}
\end{equation}
The first term represents kinetic energies of conduction electrons due to their hopping between adjacent sites where $c^\dag_{i\sigma}$ $(c_{i\sigma})$ is a creation (annihilation) operator of an electron with spin $\bm \sigma$ on the site $i$, and $t_{ij}$ denotes transfer integrals between adjacent sites $i$ and $j$. The second term represents the Kondo coupling, which describes exchange coupling between the conduction electron spins $\bm \sigma$ and the localized spins $\bm S_i$. Here $J_{\rm K}\,(> 0)$ and $\bm \sigma \equiv (\sigma_x, \sigma_y, \sigma_z)$ are the Kondo-coupling constant and a vector of the Pauli matrices, respectively. The localized spins $\bm S_i$ are treated as classical vectors, norms of which are set to be unity ($|\bm S_i|=1$). The electron filling $n_{\rm e}$ is given by,
\begin{align}
n_{\rm e} \equiv \frac{1}{2N}\sum_{i,\sigma}\braket{c_{i\sigma}^\dagger c_{i\sigma}},
\label{eq:Eq02}
\end{align}
where $N$ is the number of lattice sites.

The effects of light irradiation are considered through attaching a Peierls phase to the transfer integrals as,
\begin{align}
t_{ij} \; \rightarrow \;
t_{ij}\exp\left[ -i\bm A(\tau) \cdot (\bm r_i - \bm r_j)\right].
\label{eq:Eq03}
\end{align}
Here $\bm A(\tau)$ is a vector potential of the electromagnetic field of light, and $\tau$ and $\bm r_i$ are time and spatial coordinates of site $i$, respectively. The vector potential $\bm A(\tau)$ is time-dependent and is related with the ac electric field $\bm E(\tau)$ of light as,
\begin{align}
\bm E(\tau)=-\frac{\partial \bm A(\tau)}{\partial \tau}.
\label{eq:Eq04}
\end{align}
We adopt natural units $e=\hbar=c=1$ and set the transfer integral $t$ and the lattice constant $a$ as the units of energy and length, respectively. As a result, all the quantities in this paper are given in dimensionless forms. Table I shows unit conversions for several quantities when $t=1$ eV and $a=5$ \AA.
\begin{table}[tb]
\caption{Unit conversions for $t$=1 eV and $a$=5 \AA.}
\centering
\begin{tabular}{lll}
\hline Qunantity & Dimensionless quantity & Corresponding value \\
\hline \hline
Frequency & $\omega = \hbar\tilde{\omega}/t=1$ & $f=\tilde{\omega}/2\pi=242$ THz \\
$E$ field & $E^\omega=ea\tilde{E}^\omega/t=$1 & $\tilde{E}^\omega=20$ MV/cm \\
Time & $\tau=\tilde{\tau}t/\hbar=$1 & $\tilde{\tau}=0.66$ fs \\
\hline
\end{tabular}
\label{tab:unitconv}
\end{table}
A ground-state phase diagram of the Hamiltonian in Eq.~(\ref{eq:Eq01}) in the equilibrium case without light field, i.e., $\bm A(\tau)=0$, is shown in Fig.~\ref{Fig01}(b) as a function of the electron filling $n_{\rm e}$ when $J_{\rm K}/t=4$~\cite{Akagi2010}. In this equilibrium phase diagram, the ferromagnetic phases appear in the lower and higher filling regions of $n_{\rm e} \leq 0.22$ and $n_{\rm e} \geq 0.63$, while the 120-degree spin order appears at half filling and in the region right below it, i.e., $0.35 \leq n_{\rm e} \le 0.5$. The phase separation (PS) takes place in areas sandwiched by the ferromagnetic and 120-degree spin ordered phases. A four-sublattice all-out phase and a stripe phase also appear in narrow regions inside the PS region~\cite{Akagi2012}.

\subsection{B. Method}
We describe a numerical method used to compute real-time dynamics of the conduction electrons and the localized spins $\bm S_i$ in the Kondo-lattice model in Eq.~(\ref{eq:Eq01}). The Hamiltonian attains the time-dependence through temporal-variation of the localized spins $\{\bm S_i(\tau)\}$ as well as the time-dependent Peierls phases due to the light electromagnetic field. The time-dependent Hamiltonian $\mathcal{H}(\tau)$ satisfies the following eigenequation,
\begin{align}
\mathcal{H}(\tau)\ket{\Psi_\mu(\tau)}
=\varepsilon_\mu(\tau)\ket{\Psi_\mu(\tau)}.
\label{eq:Eq05}
\end{align}
Here $\ket{\Psi_\mu(\tau)}$ is the $\mu$th one-particle eigenstate with $\mu$ ($=1, 2,\ldots, 2N$) being a label numbering the eigenstates in ascending order with respect to the corresponding eigenenergies $\varepsilon_\mu(\tau)$. 

We simulate time evolutions of one-particle states $\ket{\tilde{\Psi}_\mu(\tau)}$ for $\mu=1, 2,\ldots, N_{\rm e}$ starting with the initial states $\ket{\tilde{\Psi}_\mu(0)}=\ket{\Psi_\mu(0)}$ by using the time-dependent Schr\"odinger equation,
\begin{align}
i\partial_\tau\ket{\tilde{\Psi}_\mu(\tau)} = \mathcal{H}(\tau)\ket{\tilde{\Psi}_\mu(\tau)},
\label{eq:Eq06}
\end{align}
where $N_{\rm e}$ is a total number of conduction electrons. We perform numerical integrations with respect to the discretized time $\tau$,
\begin{align}
\ket{\tilde{\Psi}_\mu(\tau + \Delta\tau)} &= U(\tau+\Delta\tau,\tau)\ket{\tilde{\Psi}_\mu(\tau)},
\label{eq:Eq07}
\end{align}
where the time-evolution operator $U(\tau+\Delta\tau,\tau)$ is given by~\cite{Tanaka2020,Tanaka2010},
\begin{align}
U(\tau+\Delta\tau,\tau) \equiv 
\exp\left[-i\Delta\tau\mathcal{H}\left(\tau+\frac{\Delta\tau}{2}\right)\right].
\label{eq:Eq08}
\end{align}
With this method, time evolutions of the one-particle states $\ket{\tilde{\Psi}_\mu(\tau)}$ can be calculated within sufficiently small errors of the order of $(\Delta\tau)^3$. We adopt $\Delta\tau=0.05$ in the present study, which guarantees sufficient accuracy of the simulation results. 

Real-time dynamics of the localized spins $\bm S_i(\tau)$ are simulated using the Landau-Lifshitz-Gilbert (LLG) equation,
\begin{align}
\partial_\tau\bm{S}_i= \bm{h}^{\mathrm{eff}}_i\times\bm{S}_i
+ \alpha_{\rm G}\bm{S}_i\times\partial_\tau\bm{S}_i,
\label{eq:Eq09}
\end{align}
where $\alpha_{\rm G}$ is the Gilbert-damping coefficient. The Gilbert-damping term is introduced phenomenologically to describe not only the energy dissipation of localized spins but also that of conduction electrons coupled to the localized spins via the Kondo exchange coupling. We mainly discuss the case with $\alpha_{\rm G}=0.1$ in the present paper, but the variation of its value never alters the fundamental physics argued in this paper qualitatively. We also note that the LLG equation used in the present study does not contain the Langevin-force term and thus describes the system at zero temperature. The inclusion of thermal fluctuations affects the effective exchange interactions among localized spins quantitatively, but we expect that it never affects the fundamental physics qualitatively. 

The effective local magnetic fields $\bm{h}^{\mathrm{eff}}_i$ acting on the $i$th localized spin $\bm S_i$ is given by,
\begin{align}
\bm{h}^{\mathrm{eff}}_i 
=-\left\langle \frac{\partial\mathcal{H}(\tau)}{\partial\bm{S}_i} \right\rangle 
=J_\mathrm{K}\sum_{\sigma\sigma'}
\langle c^\dag_{i\sigma}\bm{\sigma}_{\sigma\sigma'}c_{i\sigma'}\rangle.
\label{eq:Eq10}
\end{align}
This formulation is based on a mean-field decoupling of the Kondo-coupling term. This treatment is justified because the localized spins $\bm S_i$ are classical and the Kondo coupling $J_{\rm K}$ is ferromagnetic, for which the Kondo effect associated with the singlet formation between the localized spins and the conduction-electron spins is absent. We employ the Runge-Kutta method to solve this equation numerically. Here we adopt $\Delta\tau=0.05$ for a time interval of the Runge-Kutta integration for the LLG equation as well as that of the time-evolution operation for the time-dependent Schr\"odinger equation.

We should mention that the present simulations tend to be subject to an artifact, i.e., the relaxation time of localized spin dynamics varies depending on the choice of $\Delta\tau$. This artifact is specific to time-evolution simulations of the Kondo-lattice models in which the electron system and the localized spin system are coupled because their time scales are totally different. Specifically, the typical time scale of the electron system is femtoseconds, whereas that of the localized spin system is picoseconds or nanoseconds. We find that a smaller $\Delta\tau$ gives a slower relaxation, and converged behaviors are obtained when $\Delta\tau \sim 10^{-5}$. However, we have confirmed that the system is always converged to quantitatively the same nonequilibrium steady state after sufficient duration irrespective of the choice of $\Delta\tau$ or the relaxation time, and even the results of dynamical processes do not alter quantitatively when $\Delta\tau$ is as small as 0.05. Thus we consider that the artifact does not affect the present arguments on the underlying physics and mechanism.

Starting with an initial set of localized spins $\{\bm S_i(\tau=0)\}$ as well as an initial set of one-particle states $\ket{\tilde{\Psi}_\mu(\tau=0)}=\ket{\Psi_\mu(0)}$ ($\mu=1, 2, \ldots, N_{\rm e}$), we calculate time evolutions of $\ket{\tilde{\Psi}_\mu(\tau)}$ and $\{\bm S_i(\tau)\}$ by solving Eqs.~(\ref{eq:Eq06}) and (\ref{eq:Eq09}) simultaneously. The wavefunction of the conduction-electron system at time $\tau$ is given by antisymmetric Cartesian products of one-particle states as,
\begin{align}
\ket{\tilde{\Phi}(\tau)}=\ket{\tilde{\Psi}_1(\tau)} \otimes \ket{\tilde{\Psi}_2(\tau)} \otimes \cdots \otimes \ket{\tilde{\Psi}_{N_\mathrm{e}}(\tau)}.
\label{eq:Eq11}
\end{align}
Note that the one-particle states $\ket{\tilde{\Psi}_\mu(\tau)}$, which evolve according to the time-dependent Schr\"odinger equation, do not coincide with the eigenstates $\ket{\Psi_\mu(\tau)}$ of the Hamiltonian at $\tau$ in general except at $\tau=0$, and they can be expanded with the orthonormal basis set of $\ket{\Psi_\mu(\tau)}$ as,
\begin{align}
\ket{\tilde{\Psi}_\alpha(\tau)}&= \sum_{\mu = 1}^{2N}u_{\mu\alpha}(\tau)\ket{\Psi_\mu(\tau)},
\label{eq:Eq12}
\end{align}
with
\begin{align}
u_{\mu\alpha}(\tau) \equiv \braket{\Psi_\mu(\tau)|\tilde{\Psi}_\alpha(\tau)}.
\label{eq:Eq13}
\end{align}
The unitary matrix $u_{\mu\alpha}(\tau)$ satisfies $u_{\mu\alpha}(0)=\delta_{\mu\alpha}$.

We introduce the following one-body operators at $\tau$ for the electron occupation of the $\mu$th one-particle eigenstate and the total energy as,
\begin{align}
&\hat{n}_\mu(\tau)=\ket{\Psi_\mu(\tau)}\bra{\Psi_\mu(\tau)},
\label{eq:Eq14}
\\
&\hat{E}_{\rm tot}(\tau)=\frac{1}{N}\sum_{\mu=1}^{2N} \varepsilon_\mu(\tau)
\ket{\Psi_\mu(\tau)}\bra{\Psi_\mu(\tau)}.
\label{eq:Eq15}
\end{align}
Their expectation values are calculated respectively as,
\begin{align}
n_\mu(\tau)
&=\braket{\tilde{\Phi}(\tau)|\Psi_\mu(\tau)}\braket{\Psi_\mu(\tau)|\tilde{\Phi}(\tau)}
\nonumber \\
&=\sum_{\alpha=1}^{N_{\rm e}}|u_{\mu\alpha}(\tau)|^2,
\label{eq:Eq16}
\\
E_{\rm tot}(\tau)
&=\frac{1}{N}\sum_{\alpha=1}^{N_{\rm e}} \sum_{\mu=1}^{2N} \varepsilon_\mu(\tau)
|u_{\mu\alpha}(\tau)|^2.
\label{eq:Eq17}
\end{align}
The expectation value of the local electron spin density $\bm \sigma_i(\tau)$, which appears in Eq.~(\ref{eq:Eq10}), can be calculated as,
\begin{align}
\bm \sigma_i(\tau)
&=\braket{\tilde{\Phi}(\tau)|c^\dag_{i\sigma}
\bm \sigma_{\sigma\sigma'}
c_{i\sigma'}|\tilde{\Phi}(\tau)}
\nonumber \\
&=\sum_{\alpha=1}^{N_{\rm e}} \bm \sigma_{\sigma\sigma'} 
\braket{\tilde{\Psi}_\alpha(\tau)|i\sigma}
\braket{i\sigma'|\tilde{\Psi}_\alpha(\tau)}.
\label{eq:Eq18}
\end{align}

To characterize the magnetization structure and its time evolution, we calculate several quantities associated with the localized spins $\{\bm S_i(\tau)\}$, i.e., the spin structure factors in the momentum space $\hat{S}(\bm q,\tau)$, the local spin chirality vectors $\bm C_i(\tau)$, and the local vorticities $\Omega_i(\tau)$ associated with the vectors $\bm C_i(\tau)$. In the following, we write these time-dependent quantities by omitting the time variable $\tau$ for simplicity. The spin structure factor is given by,
\begin{align}
\hat{S}(\bm{q})=\frac{1}{N^2}\sum_{i,j}(\bm{S}_i\cdot\bm{S}_j) 
e^{i\bm{q}\cdot(\bm{r}_i - \bm{r}_j)}.
\label{eq:Eq19}
\end{align}
The spin chirality vector $\bm C_i$ at site $i$ is defined with three localized spins $\bm S_i$, $\bm S_{i+\hat{a}}$ and $\bm S_{i+\hat{a}+\hat{b}}$ sharing an up-pointing triangular plaquette as,
\begin{align}
\bm{C}_{i}=\frac{2}{3\sqrt{3}}
(\bm{S}_i\times\bm{S}_{i+\hat{a}}
+\bm{S}_{i+\hat{a}}\times\bm{S}_{i+\hat{a}+\hat{b}}
+\bm{S}_{i+\hat{a}+\hat{b}}\times\bm{S}_i).
\label{eq:Eq20}
\end{align}	
For definitions of the primitive translational vectors $\bm a$ and $\bm b$, see Fig.~\ref{Fig02}. The normalization factor is introduced such that the norm $|\bm C_i|$ becomes unity for a perfectly coplanar 120-degree alignment of the three localized spins. The averaged vector spin chirality $C$ is given by,
\begin{align}
C=\frac{1}{N}\sum_i |\bm C_i|.
\label{eq:Eq21}
\end{align}	
We also define the local vorticity $\Omega_i$ at site $i$ as,
\begin{align}
\Omega_i=\sum_{j=1}^{6}\cos^{-1}
\left[\frac{\bm{C}_j\cdot\bm{C}_{j+1}}{|\bm{C}_j||\bm{C}_{j+1}|}\right]
\mathrm{sgn}\left[(\bm{C}_j\times\bm{C}_{j+1})_z\right],
\label{eq:Eq22}
\end{align}
where $\bm{C}_j$ ($1\le j \le 7$, $\bm{C}_7 \equiv \bm{C}_1$) are the spin chirality vectors at six sites adjacent to the site $i$ numbered in a counterclockwise order, and $\mathrm{sgn}(x)$ is a sign function. This quantity takes $+2\pi$ at a vortex core and $-2\pi$ at an antivortex core.

We use a simple triangular lattice of $N=L^2$ sites and impose periodic boundary conditions. In this paper, we mainly discuss the results obtained for $L=18$. However, we have confirmed that predicted phenomena and proposed physical mechanisms do not alter qualitatively even in larger-sized systems by performing the calculations for various system sizes up to $L=60$ with several sets of parameters. We start with a ground-state spin configuration as an initial state. In most cases, we examine the parameter ranges, for which the ground state is ferromagnetic. We adopt a list-mapping technique to reduce computational costs, because the Hamiltonian matrix used in this study is sparse, and only a small number of components have nonzero values. 

When we irradiate the ferromagnetic state with light, a special treatment is required for the simulations. The perfectly polarized ferromagnetic state is energetically stable even under the photoirradiation and is hardly changed to other magnetic states by simple uniform irradiation with light because the effective magnetic field $\bm h_i^{\rm eff}$ in the LLG equation is spatially uniform even in the presence of the light electromagnetic field. In real systems, there are several factors which disturb the perfect ferromagnetic polarization or the uniformity of magnetization alignment such as defects, impurities, thermal fluctuations, and edges of the samples. These factors trigger nucleation of tiny spin-flipped or spin-rotated areas under the photoirradiation, and growth of these nucleated areas results in the photoinduced magnetic phase transition. To mimic such a situation in real systems, we introduce a small amount of randomness in orientations of the ferromagnetic magnetizations in the initial state.
\section{RESULTS}
\subsection{A. Real-time dynamics}
\begin{figure}[tb]
\centering
\includegraphics[scale=0.5]{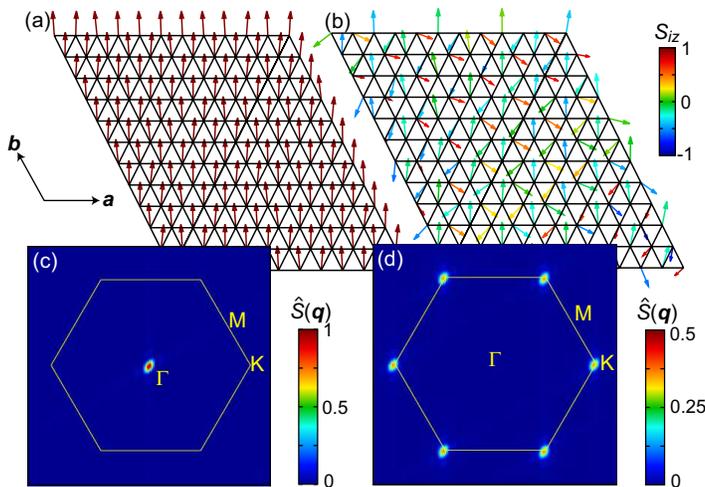}
\caption{(a), (b) Typical real-space configurations of the localized spins $\{\bm S_i\}$ for (a) the ferromagnetic state before photoirradiation and (b) the 120-degree spin ordered state in the photodriven system. (c), (d) Spin structure factors $\hat{S}(\bm q)$ in the momentum space for these magnetic states. The calculations are performed for the Kondo-lattice model with $J_{\rm K}/t=4$ and $n_{\rm e}=1/6$ irradiated by circularly polarized light of $E^\omega=1.3$ and $\omega=0.8$. A triangular lattice with $18\times18$ sites is used for the calculations.}
\label{Fig02}
\end{figure}
We first discuss results on the photoinduced magnetic phase and the real-time dynamics of the photoinduced magnetic phase transition. We examine a system irradiated with circularly polarized light by considering the following form for time-dependent vector potential $\bm A(\tau)$ in Eq.~(\ref{eq:Eq03}),
\begin{align}
\bm A(\tau)=\frac{E^\omega}{\omega}\left(\cos(\omega\tau), \sin(\omega\tau)\right).
\label{eq:Eq23}
\end{align}
Here $E^\omega$ is the amplitude of ac electric field of light. 

In Fig.~\ref{Fig02}, we show the localized spin structure before the photoirradiation and that during the photoirradiation after sufficient duration. The spatial configurations of localized spins at (a) $\tau = 0$ and (b) $\tau = 2000$ show that the initial ferromagnetic order is changed to the 120-degree spin order by the photoirradiation. Here the localized spins $\{\bm S_i\}$ in the ferromagnetic state are all oriented perpendicular to the plane, and we draw the magnetization vectors $\bm S_i=(S_{ix}, S_{iy}, S_{iz})$ in Fig.~\ref{Fig02}(a) by arrows $(S_{ix}, S_{iz})$ lying in the plane for clear visuality. This photoinduced change of the magnetic structure is also clearly seen in the reciprocal space. The initial spin structure factor before the photoirradiation (at $\tau=0$) in Fig.~\ref{Fig02}(c) has a sharp single peak at $\Gamma$ point, while that during the photoirradiation at $\tau=2000$ in Fig.~\ref{Fig02}(d) has peaks at K points on the hexagonal first Brillouin zone. In the following, we focus on the magnitudes of spin structure factors at $\Gamma$ and K points, $\hat{S}(\Gamma)$ and $\hat{S}({\rm K})$, as measures of respective spin correlations, and use their difference $\Delta\hat{S} \equiv \hat{S}(\Gamma)-\hat{S}({\rm K})$ as an indicator of the dominant spin correlation.

\begin{figure}[tb]
\centering
\includegraphics[scale=0.5]{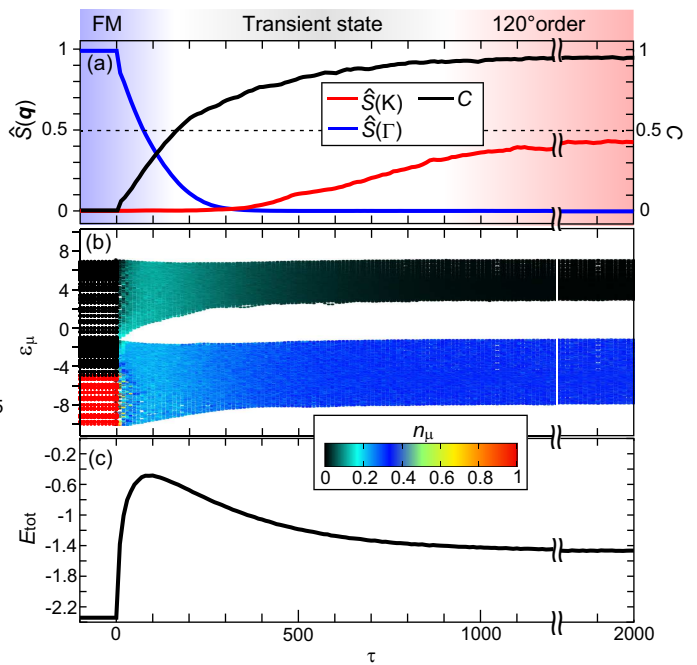}
\caption{(a) Time profiles of the spin structure factors $\hat{S}(\bm q)$ at $\Gamma$ and K points in the first Brillouin zone and the averaged vector spin chirality $C=\sum_i |\bm C_i|/N$ for the photoinduced phase transition from ferromagnetic (FM) to 120-degree spin ordered states. The quantity $C$ is normalized to be unity for a perfect (single-domain) 120-degree spin order. In the simulation, however, it converges to a slightly smaller value of $\sim$ 0.95 even after sufficient duration because of domain formation as discussed later. (b) Time profiles of the eigenenergies $\varepsilon_\mu$ and the electron occupations $n_\mu$. (c) Time profile of the total energy $E_{\rm tot}$. The calculations are performed for the Kondo-lattice model with $J_{\rm K}/t=4$ and $n_{\rm e}=1/6$ irradiated by circularly polarized light of $E^\omega=1.3$ and $\omega=0.8$. A triangular lattice with $18\times18$ sites is used, and the Gilbert-damping coefficient is set to be $\alpha_{\rm G}=0.1$ for the calculations.}
\label{Fig03}
\end{figure}
Figure~\ref{Fig03} shows simulated time profiles of several physical quantities. In the time profiles of $\hat{S}(\Gamma)$ and $\hat{S}({\rm K})$ in Fig.~\ref{Fig03}(a), we find that the ferromagnetic correlation manifested by $\hat{S}(\Gamma)$ abruptly decreases and vanishes immediately after starting the photoirradiation, whereas the 120-degree spin correlation manifested by $\hat{S}({\rm K})$ grows gradually. The averaged spin chirality $C$ in Eq.~(\ref{eq:Eq21}) provides another measure of the 120-degree spin correlation. The simulated time profile of $C$ in Fig.~\ref{Fig03}(a) shows a monotonic increase along with the growth of 120-degree spin correlation. This quantity should be unity in the perfect 120-degree spin order of single domain, but it converges to a slightly smaller value of $\sim$ 0.95 even after sufficient duration. This reduction is attributable to formation of domains with different 120-degree ordered planes as discussed later. 

We also find that the spin chirality $C$ grows rapidly accompanied by destruction of the ferromagnetic order, which is considerably quick as compared with the growth of spin structure factor $\hat{S}({\rm K})$. This indicates that a lot of small domains with chiral spin configurations are locally created in the initial process right after starting the photoirradiation, and the 120-degree spin configurations are formed through subsequent alignment of the local spin chiralities. Note that both $\hat{S}(\Gamma)$ and $\hat{S}({\rm K})$ are suppressed to be zero at $\tau \sim 300$. In a snapshot of the spatial spin configuration at this peculiar moment (not shown), we find that the spins align not fully at random, but they seem to evolve towards the 120-degree order. Namely, if we look at each triangular plaquette, three spins form nearly 120-degree alignment at some plaquettes, and two of three spins align with an angle of 120$^\circ$ at some other plaquettes. These local (imperfect) 120-degree plaquettes give rise to the finite spin chirality $C$, but they do not have long-range correlations, resulting in the vanishing spin correlations ($\hat{S}(\Gamma)=\hat{S}({\rm K})=0$) at this moment.

To clarify a microscopic mechanism of this photoinduced magnetic phase transition, we then study the time evolution of conduction electrons under photoirradiation. Figure~\ref{Fig03}(b) shows time profiles of eigenenergies $\varepsilon_\mu$ and the number of electrons $n_\mu$ that occupy the corresponding eigenstates of the Hamiltonian at time $\tau$, where $\mu$ is an index of the eigenstates. We find that the initial band structure before photoirradiation is gapless, but a gap starts opening and grows after starting the photoirradiation. The photoinduced gap separates the originally gapless continuum band into lower and upper portions. This gap formation is attributable to the Kondo exchange coupling $J_{\rm K}$ and the dynamical localization effect~\cite{Dunlap1986,Grossmann1991,Ishikawa2014,Holthaus1992,Kayanuma2008,Eckardt2005,Lignier2007,Ohmura2021}, i.e., the bandwidth narrowing in driven tight-binding systems. According to the Floquet theory for photoirradiated tight-binding models, transfer integrals $t_{ij}$, to which the Peierls phases associated with the time-dependent vector potential $\bm A(\tau)$ are attached, are renormalized by a factor of the Bessel function $J_0(\mathcal{A}_{ij})$ where $\mathcal{A}_{ij}=e{\bm E}^\omega \cdot (\bm r_i-\bm r_j)/\hbar\omega$~\cite{Yonemitsu2017}. Before the photoirradiation, the bandwidth exceeds the magnitude of the exchange-splitting due to the Kondo coupling, resulting in the gapless band. On the other hand, the photoirradiation suppresses the bandwidth through the dynamical localization effect, and eventually the gap opens when the magnitude of exchange splitting dominates the bandwidth.

A time profile of the total energy $E_{\rm tot}$ of conduction electrons in Fig.~\ref{Fig03}(c) shows an abrupt increase immediately after starting the photoirradiation and takes a maximum. This abrupt increase is caused by the photoexcitation of conduction electrons and resulting electron occupations of higher-lying eigenstates. After taking a maximum, the total energy starts decreasing and converges to a certain value through relaxation of the electron distribution after the band gap sufficiently grows. These behaviors suggest that a process of the dynamical magnetic phase transition contains three important stages, i.e., the photoexcitation of conduction electrons, the photoinduced band-gap opening, and the subsequent redistribution of excited electrons through relaxation.

\begin{figure*}[tbh]
\centering
\includegraphics[scale=0.5]{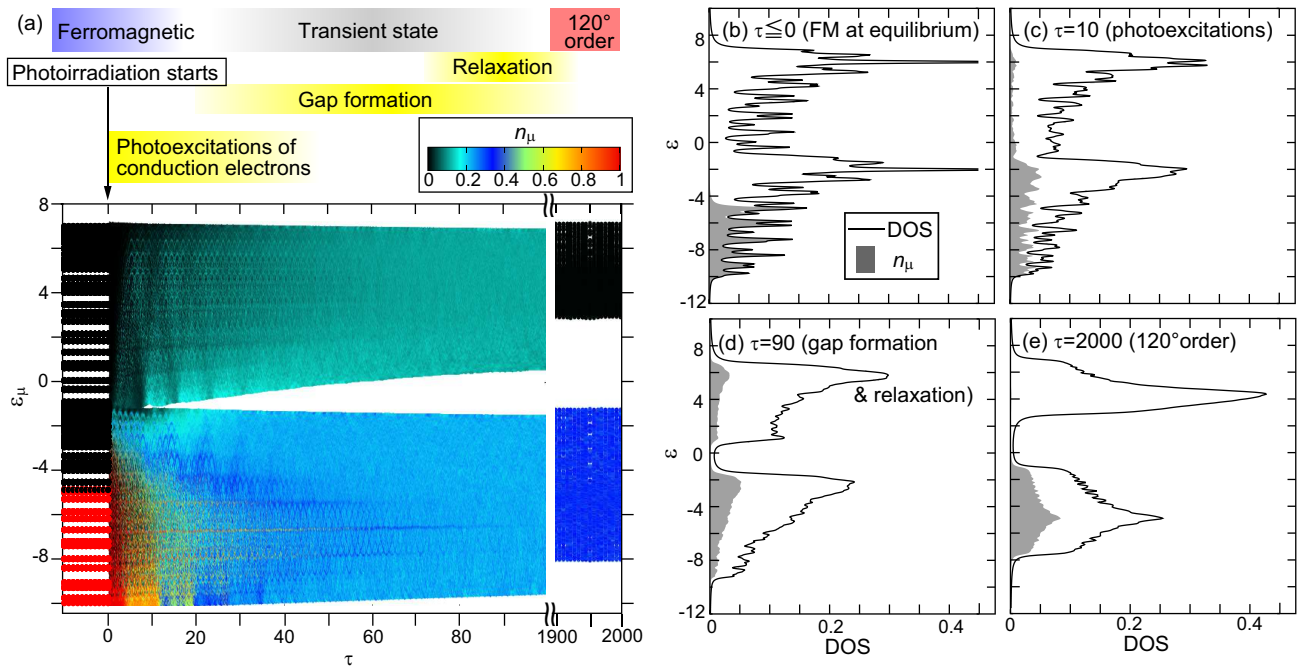}
\caption{(a) Time profiles of the eigenenergies $\varepsilon_\mu$ and the electron occupations $n_\mu$ of the corresponding eigenstates in the initial process and at the final stage of the photoinduced dynamical phase transition, which magnifies a time range of $0 \le \tau \le 100$ and that around $\tau \sim 2000$ in Fig.~\ref{Fig03}(b). (b)-(e) Calculated densities of states (DOS) and electron occupations $n_\mu$ at various stages of the photoinduced phase transition. (b) those in the ferromagnetic state before the photoirradiation ($\tau \leq 0$), (c) those during the photoexcitation of conduction electrons immediately after starting the irradiation ($\tau=10$), (d) those during the band-gap formation and relaxation in the transient process ($\tau=90$), and (e) those in the nonequilibrium steady phase of the 120-degree spin order after sufficient duration ($\tau=2000$). The shaded areas indicate the electron-occupation rate of the bands. The parameters and conditions used for the calculations are the same as those for Fig.~\ref{Fig03}.}
\label{Fig04}
\end{figure*}

These stages can be clearly seen in the simulated time evolution of electronic structures, i.e., the band structure, the density of states, and the electron occupations. Figure~\ref{Fig04}(a) shows time profiles of the eigenenergies $\varepsilon_\mu$ and the electron occupations $n_\mu$ in the initial process and at the final stage, which magnifies a time range of $0 \leq \tau \leq 100$ and that around $\tau \sim 2000$ in Fig.~\ref{Fig03}(b). Before the photoirradiation ($\tau=0$), the system has a gapless continuum band which spreads over an energy range of $-10 \leq \varepsilon_\mu \leq 7$, and only the eigenstates at lower energies ranging in $-10 \leq \varepsilon_\mu \leq -4$ are occupied by electrons. These aspects can also be seen in the density of states and the electron occupations $n_\mu$ at $\tau=0$ in Fig.~\ref{Fig04}(b).

Immediately after starting the photoirradiation, the conduction electrons, which originally occupy the lower-lying eigenstates, are excited to the higher-lying states within the gapless continuum band. Indeed the originally black-colored higher-lying eigenstates become slightly brighter or light-blue-colored in the time range of $0 \leq \tau \leq 15$ in Fig.~\ref{Fig04}(a), while the band is still gapless in this very initial stage. The density of states and the electron occupations at $\tau=10$ in Fig.~\ref{Fig04}(c) clearly show the gapless continuum band and the occupations of higher-lying eigenstates by photoexcited electrons. From $\tau\sim 15$, a band gap starts opening, and its magnitude gradually increases as time proceeds, which results in the formation of upper and lower bands separated by the gap. Through this gapped-band formation, the excited electrons in the higher-lying states are left in the upper band. Moreover, while the band gap is small, the conduction electrons are continued to be excited from the lower band to the upper band. The numbers of electrons in the lower-energy (higher-energy) states gradually decrease (increase) in this initial process. Consequently, the excited electrons become to be distributed to all the eigenstates in the lower band nearly uniformly as indicated by the uniformly blue-colored lower band after $\tau\sim 40$ in Fig.~\ref{Fig04}(a). Moreover, even the eigenstates in the upper band become to be occupied by the photoexcited electrons nearly uniformly as indicated by the uniformly light-blue-colored upper band after $\tau\sim 40$. These aspects can also be seen in the density of states and the electron occupations at $\tau=90$ in Fig.~\ref{Fig04}(d).

As we continue the photoirradiation, the widths of both upper and lower bands become more and more suppressed, which leads to a further growth of band gap. After the band gap sufficiently grows, the photoexcitation of conduction electrons from the lower band to the upper band can no longer happen. Alternatively, the conduction electrons that occupy the upper band start falling to the lower band through dissipation. In the present simulations, the effect of energy dissipation is taken into account by introducing the Gilbert-damping term in the LLG equation. The excited electrons lose their energies and fall to the lower-lying states through coupling to the dissipative dynamics of localized spins via the Kondo coupling. Consequently, the electron occupations of the upper band vanish, while the nearly uniform electron occupations of the lower band are realized after sufficient duration [see Fig.~\ref{Fig04}(e)]. Importantly, this electronic structure resembles that in the half-filled system at equilibrium, in which the static 120-degree spin order is stabilized. At equilibrium, the Fermi-surface nesting favors the 120-degree spin order in the present triangular Kondo-lattice system near the half filling~\cite{Akagi2010,Akagi2012}. When the eigenstates in the lower band are nearly uniformly occupied, the electrons that occupy the higher-lying states in the lower band are subject to the similar nesting vectors in this pseudo half-filled situation. This situation is expected to stabilize the 120-degree spin order in the present photoirradiated nonequilibrium system.

It is worth mentioning that both upper and lower bands are partially occupied by conduction electrons in the transient state [see Fig.~\ref{Fig04}(d)]. We expect that the ferromagnetic correlation might be caused by the photoexcited electrons in the upper band in this photodriven system. According to the phase diagram in Fig.~\ref{Fig01}(b), the ferromagnetic phase appears at higher filling of $n_{\rm e} \geq 0.625$ in the equilibrium case. This equilibrium ferromagnetic phase is stabilized by the double-exchange mechanism~\cite{Zener1951,Anderson1955,Gennes1960}, which favors a ferromagnetic alignment of the localized spins to maximize a gain of the kinetic energies of the conduction electrons coupled to the localized spins. We expect that this mechanism also works in the transient state under the photoirradiation. Indeed, the ferromagnetic correlation or the amplitude of $\hat{S}(\Gamma)$ vanishes at $\tau\sim 300$ when the electron occupation of the upper band is significantly suppressed by the fully opened band gap.

\subsection{B. Domains and $Z_2$ vortices}
\begin{figure*}[p]
\centering
\includegraphics[scale=1.0]{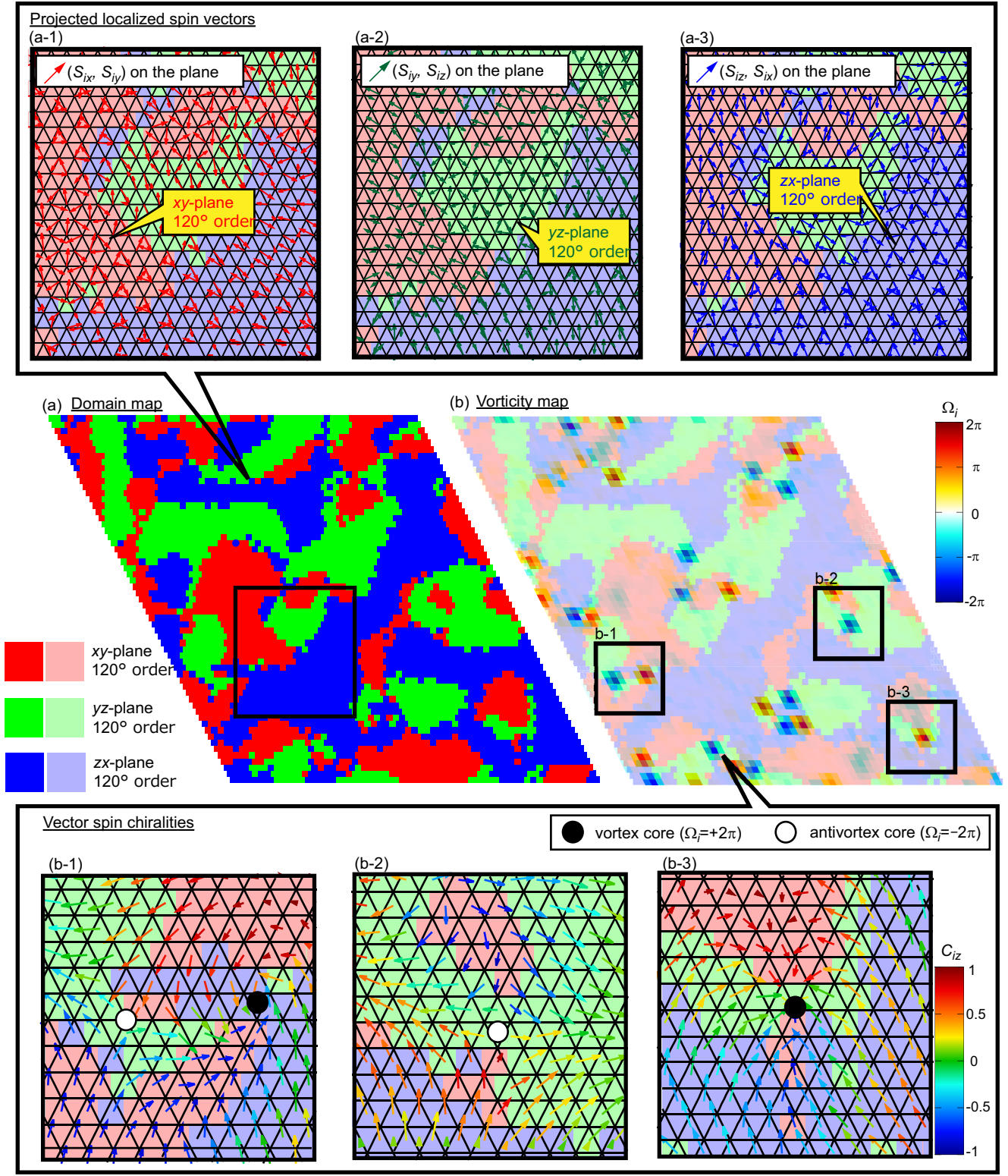}
\caption{(a) Snapshot of the spatial configuration of three kinds of domains associated with the 120-degree spin orders in the $xy$, $yz$, and $zx$ planes. This kind of spatial pattern of domains emerges in the photoinduced nonequilibrium 120-degree spin ordered phase. The area indicated by a solid square is magnified in three panels above, where projected localized spin vectors, (a-1) $(S_{ix}, S_{iy})$, (a-2) $(S_{iy}, S_{iz})$, and (a-3) $(S_{iz}, S_{ix})$, are presented by arrows on the $xy$ plane. In each panel, the spin vectors form coplanar 120-degree orders within corresponding domains. (b) Snapshot of the spatial configuration of local vorticities $\Omega_i$ associated with the $Z_2$ vortices and antivortices composed of the local spin chirality vectors $\bm C_i$. This quantity takes $+2\pi$ ($-2\pi$) at a core of vortex (antivortex). The spatial map of 120-degree ordered domains is also presented by light colors for reference, which shows that the vortices and antivortices appear at crossing points of domain boundaries or points at which several domains meet. Some of the vortices or antivortices are magnified in panels under this figure, in which local spin chirality vectors $\bm C_i$ are presented by arrows and colors. The calculations are performed for $J_{\rm K}/t=4$, $n_{\rm e}=1/6$, $E^\omega=1.3$, $\omega=0.8$, and $\alpha_{\rm G}=1$ using a triangular lattice of $60\times60$ sites. The snapshot is taken after sufficient duration at $\tau$=2000.}
\label{Fig05}
\end{figure*}
The averaged spin chirality $C$ in Eq.~(\ref{eq:Eq21}) should be unity for a single-domain 120-degree spin ordered phase. In our simulations, however, this quantity converges to a slightly smaller value of $\sim 0.95$ as seen in Fig.~\ref{Fig03}(a). This slight reduction is attributable to formation of domains with different 120-degree ordered planes. When magnetic anisotropies are absent or sufficiently weak, the 120-degree spin order has infinite degeneracy with respect to the choice of ordering planes. Namely, the three sublattice spins forming the 120-degree order can align within any planes in magnetically isotropic systems. In the photoinduced nonequilibrium steady phase, the 120-degree ordered domains with different ordering planes should appear. To study such a domain formation in the dynamical phase, we perform numerical simulations for a larger-sized system. Although the ordering plane is a continuous degree of freedom, we classify the local ordering planes into three kinds for simplicity, i.e., $xy$, $yz$, and $zx$ planes according to the largest component of the local spin chirality vectors $\bm C_i=(C_{ix}, C_{iy}, C_{iz})$. Specifically, the local ordering planes are classified into $xy$, $yz$, and $zx$ when the largest component is $C_{iz}$, $C_{ix}$, and $C_{iy}$, respectively.

Figure~\ref{Fig05}(a) shows a snapshot of the spatial configuration of three kinds of domains in the photoinduced 120-degree spin ordered phase. The simulation was performed by taking a system of $60\times60$ sites with $J_{\rm K}/t=4$, $n_{\rm e}=1/6$, and $\alpha_{\rm G}=1$ irradiated by circularly polarized light with $E^\omega=1.3$ and $\omega=0.8$. This snapshot is taken after sufficient duration at $\tau=2000$. The area indicated by a solid square in this figure is magnified in three panels above, i.e., (a-1), (a-2), and (a-3). In each panel, projected localized spin vectors, i.e., (a-1) $(S_{ix}, S_{iy})$, (a-2) $(S_{iy}, S_{iz})$, and (a-3) $(S_{iz}, S_{ix})$, are presented by arrows. We find that the spin vectors $\{\bm S_i\}$ in each domain form a coplanar 120-degree spin order within the specific ordering plane.

In this spatial domain pattern, we naively expect that the alignment of spin chirality vectors $\bm C_i$ are twisted at domain boundaries because the vectors $\bm C_i$ are pointed in different directions between the domains. Subsequently, it is anticipated that the twisted spin chirality vectors $\bm C_i$ form vortex or antivortex structures at points where multiple domains meet, which are referred to as $Z_2$ (anti)vortex~\cite{Kawamura1984,Kawamura2010,Okubo2010,Aoyama2020,Tomiyasu2021}. In Fig.~\ref{Fig05}(b), we present a snapshot of the spatial configuration of local vorticities $\Omega_i$ associated with the $Z_2$ vortices and antivortices composed of the local spin chirality vectors $\bm C_i$. This quantity takes $+2\pi$ ($-2\pi$) at a core of the vortex (antivortex). The spatial map of domains is also presented by light colors for reference, which shows that the vortices and antivortices appear at crossing points of multiple domain boundaries. Some of the vortices and antivortices are magnified in panels below the figure, in which the local spin chirality vectors $\bm C_i$ are presented by arrows and colors.

\section{CONCLUSION}
In this paper, we have theoretically proposed a possible photoinduced magnetic phase transition to the nonequilibrium 120-degree spin ordered phase in the Kondo-lattice model on a triangular lattice by numerically simulating the spatiotemporal dynamics of the conduction electrons and the localized spins under photoirradiation. It has turned out that the process of this photoinduced magnetic phase transition contains three important stages, that is, the photoexcitation of conduction electrons, the band-gap formation due to the dynamical localization effect, and the relaxation of the excited conduction electrons through dissipations. Immediately after starting the photoirradiation, the conduction electrons are once excited to higher-lying states within the originally gapless continuum band in the Kondo-lattice system away from the half-filing. Subsequently, the bandwidth becomes suppressed because of the dynamical localization effect under the photoirradiation. When the exchange splitting energy due to the Kondo coupling dominates the reduced bandwidth, a gap dividing the continuum band into upper and lower portions is formed. The magnitude of gap becomes larger as the bandwidth becomes narrower. After this band reconstruction with a sufficiently grown band gap, the excited conduction electrons in the upper band become relaxed to fall to the lower band through the energy dissipation. Eventually, the conduction electrons are distributed to almost all the eigenstates constituting the lower band nearly uniformly in the nonequilibrium steady states of the photodriven system. This electronic structure resembles that in the half-filled system at equilibrium. In the Kondo-lattice model, ordering of the localized spins at equilibrium is governed by nesting vectors of the Fermi surface determined by the electron filling. The conduction electrons in the photodriven system with this pseudo half-filled electronic structure are subject to the similar nesting vectors and thus can stabilize the 120-degree spin order as those in the half-filled system at equilibrium do.

We have also discussed formation of domains with different 120-degree ordered planes and the emergence of topological vortices and antivortices composed of local vector spin chiralities (i.e., the $Z_2$ vortices) in the photoinduced nonequilibrium steady phase of the 120-degree spin order. 
In fact, we have also investigated the dependencies of the photoinduced phase transition on the light parameters (i.e., amplitude, frequency, and polarization), the electron filling, the strength of Kondo coupling, and the antiferromagnetic exchange coupling among the localized spins to discuss favorable conditions and situations to observe the predicted phenomenon (see Appendices). Through these investigations, we have found that the predicted phenomenon can be realized by a relatively weak light intensity particularly in the presence of the antiferromagnetic exchange coupling, and thus the experimental observation is feasible. We have also found that the phenomenon is induced by a laser light irrespective of its polarization. So far, only limited magnetic structures have been argued to emerge by photoirradiation, e.g., simple collinear (anti)ferromagnetic orders, local magnetic defects, and paramagnetic states. The present work is expected to pave a way to optical creation of complex noncollinear magnetism as a nonequilibrium steady phase in the photodriven system.

\section{Appendices}
In the following appendices, we discuss several factors that might affect the proposed photoinduced phase transition to the 120-degree spin order. We investigate dependencies of this phenomenon on the strength of Kondo coupling $J_{\rm K}$, the electron filling $n_{\rm e}$, and the light polarization. We also discuss the effect of antiferromagnetic exchange coupling $J_{\rm AFM}$ among the localized spins, the influence of the variation of Gilbert-damping coefficient, and feasibility of the experimental realization. To see whether the photoinduced magnetic phase transition occurs, we focus on the spin structure factors at $\Gamma$ and K points in the first Brillouin zone, $\hat{S}(\Gamma)$ and $\hat{S}({\rm K})$, after sufficient photoirradiation, i.e., at $\tau=2000$. These quantities represent strengths of ferromagnetic and 120-degree spin correlations, respectively. Note that $\hat{S}(\Gamma)$ becomes unity for a perfect ferromagnetic state, whereas $\hat{S}({\rm K})$ becomes one-half for a perfect single-domain 120-degree spin ordered state . The difference $\Delta \hat{S} \equiv \hat{S}({\rm K})-\hat{S}(\Gamma)$ is a measure of the dominant spin correlation under photoirradiation. 

\subsection{Appendix A. Kondo coupling}
\begin{figure*}[tb]
\centering
\includegraphics[scale=1.0]{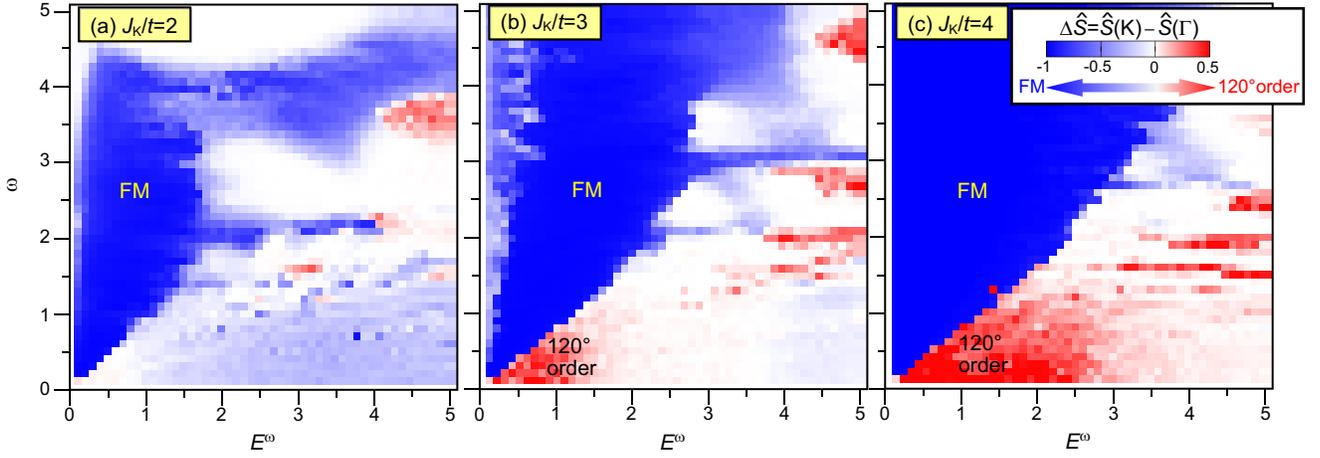}
\caption{Color maps of the relative spin correlation $\Delta \hat{S} \equiv \hat{S}({\rm K})-\hat{S}(\Gamma)$ for the photoirradiated Kondo-lattice model on a triangular lattice after sufficient duration, i.e., at $\tau=2000$ in plane of the amplitude $E^\omega$ and the frequency $\omega$ of circularly polarized light for various strengths of the Kondo coupling, i.e., (a) $J_{\rm K}/t=2$, (b) $J_{\rm K}/t=3$, and (c) $J_{\rm K}/t=4$. The calculations are performed for a system of $18 \times 18$ sites with periodic boundary conditions by setting $n_{\rm e}=1/6$ and $\alpha_{\rm G}=1$.}
\label{Fig06}
\end{figure*}
We first examine dependency on the strength of Kondo coupling $J_{\rm K}$. Figure~\ref{Fig06} presents color maps of the relative spin correlation $\Delta \hat{S}$ for the photodriven systems in plane of the amplitude $E^\omega$ and the frequency $\omega$ of circularly polarized light for various strengths of $J_{\rm K}$, i.e., (a) $J_{\rm K}/t=2$, (b) $J_{\rm K}/t=3$, and (c) $J_{\rm K}/t=4$. The calculations are performed for the Kondo-lattice model with $n_{\rm e}=1/6$ on a triangular lattice of $18 \times 18$ sites with periodic boundary conditions. We adopt a rather large Gilbert-damping coefficient of $\alpha_{\rm G}=1$ to achieve quick convergence for saving the computational time. We have confirmed for some parameter sets that the final magnetic states do not alter even if we slowly relax the system with a small Gilbert-damping coefficient of $\alpha_{\rm G}=0.1$. 

We observe clear photoinduced magnetic phase transition to the nonequilibrium 120-degree spin ordered state when the Kondo coupling is rather strong as $J_{\rm K}/t=3$ and $J_{\rm K}/t=4$ indicated by positive $\Delta \hat{S}$, while the phase transition is obscure for a weaker coupling of $J_{\rm K}/t=2$. This indicates that a stronger Kondo coupling is favorable for the photoinduced 120-degree spin order. This tendency can be understood as follows. The Kondo coupling $J_{\rm K}$ governs a magnitude of the band gap which separates the upper and lower bands in the photoirradiated system. According to the literature, the 120-degree spin order is stabilized near the half filling in the equilibrium case~\cite{Akagi2010,Akagi2012}. Under the photoirradiation, the electronic structure resembling the half-filled system is realized as a nearly uniform distribution of the photoexcited electrons in the lower band well-separated from the upper band by the gap. A large band gap induced by a strong Kondo coupling is important for realizing such an electronic structure.

\subsection{Appendix B. Electron filling}
\begin{figure*}[tb]
\centering
\includegraphics[scale=1.0]{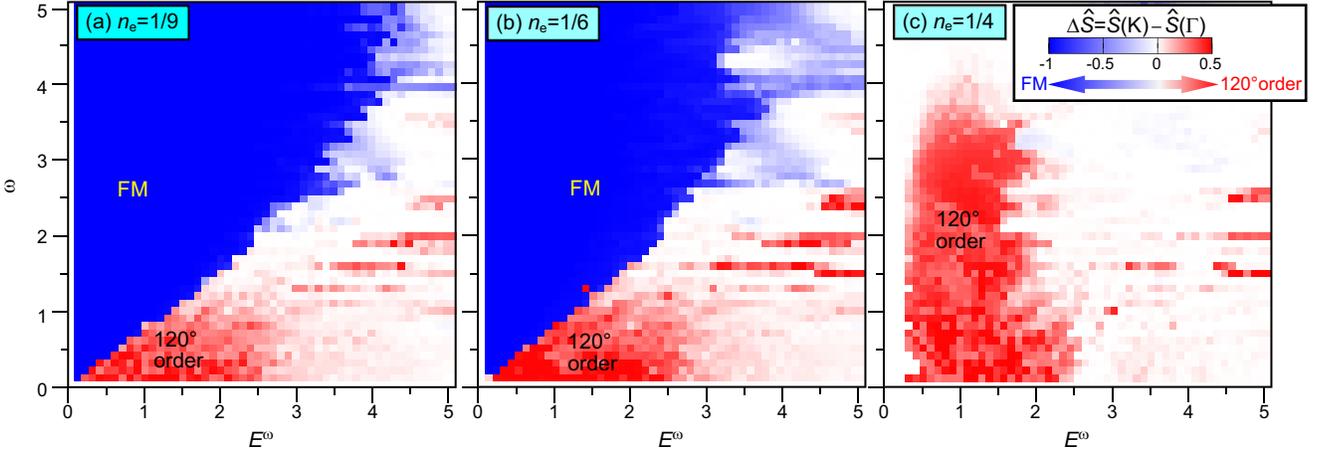}
\caption{Color maps of the relative spin correlation $\Delta \hat{S} \equiv \hat{S}({\rm K})-\hat{S}(\Gamma)$ for the photoirradiated Kondo-lattice model on a triangular lattice after sufficient duration, i.e., at $\tau=2000$, in plane of the amplitude $E^\omega$ and the frequency $\omega$ of circularly polarized light for various electron fillings, i.e., (a) $n_{\rm e}=1/9$, (b) $n_{\rm e}=1/6$, and (c) $n_{\rm e}=1/4$. The calculations are performed for a system of $18 \times 18$ sites with periodic boundary conditions by setting $J_{\rm K}/t=4$ and $\alpha_{\rm G}=1$. Surprisingly, the photoinduced magnetic transition to the 120-degree spin ordered state occurs even for a lower electron filling of $n_{\rm e}=1/9$ away from the half filling. In the case of $n_{\rm e}=1/4$, we perform the calculations starting with a four-sublattice all-out state as an initial state instead of the ferromagnetic state according to a theoretically predicted ground-state spin configuration at this electron filling. Figure (c) shows that the initial four-sublattice state disappears and the 120-degree spin ordered state emerges in a wide area over the $E^\omega$-$\omega$ plane.}
\label{Fig07}
\end{figure*}
We next study dependency on the electron filling $n_{\rm e}$ by constructing color maps of the relative spin correlation $\Delta \hat{S}$ for the photodriven systems in plane of the amplitude $E^\omega$ and the frequency $\omega$ of circularly polarized light for various values of the electron filling $n_{\rm e}$, i.e., (a) $n_{\rm e}=1/9$, (b) $n_{\rm e}=1/6$, and (c) $n_{\rm e}=1/4$ as presented in Fig.~\ref{Fig07}. The calculations are performed for the Kondo-lattice model with $J_{\rm K}/t=4$ on a triangular lattice of $18 \times 18$ sites with periodic boundary conditions by setting $\alpha_{\rm G}=1$. When the electron filling is rather small as $n_{\rm e}=1/9$ and $n_{\rm e}=1/6$, the photoinduced magnetic phase transition to the nonequilibrium 120-degree spin ordered state occurs only when $E^\omega \gtrsim \omega$ [see Figs.~\ref{Fig07}(a) and (b)]. On the contrary, for a large electron filling of $n_{\rm e}=1/4$, the 120-degree spin ordered state appears in a wide area, which spreads even to the region of $E^\omega \lesssim \omega$ [see Fig.~\ref{Fig07}(c)]. These results indicate that the higher electron filling is favorable for the photoinduced 120-degree spin ordered state, but it can emerge even at lower electron fillings when the light amplitude $E^\omega \equiv ea\tilde{E}^\omega/t$ is larger than the light frequency $\omega \equiv \hbar\tilde{\omega}/t$ with $\tilde{E}^\omega$ and $\tilde{\omega}$ being the dimensionfull amplitude and frequency of light.

In Figs.~\ref{Fig07}(a) and (b), the clear 120-degree spin ordered state with positive $\Delta \hat{S}$ appears in the frequency region of $\omega \lesssim 3$, whereas $\Delta \hat{S}$ tends to be suppressed in the high frequency region of $\omega \gtrsim 3$, indicating that the high-frequency light cannot necessarily induce the 120-degree spin order efficiently. This tendency can be understood by considering the band gap due to the Kondo coupling. For the present calculations using $J_{\rm K}/t=4$, the magnitude of the band gap is $\sim 3$ in the photoirradiated nonequilibrium system [see Fig.~\ref{Fig04}(e)]. When the light frequency $\omega$ exceeds this value, the conduction electrons in the lower band can be excited to the upper band through crossing the band gap even in the nonequilibrium steady state. The resulting electron occupations of the upper band work destructively for the formation of pseudo half-filled electronic structure required for the emergence of 120-degree spin ordered state. Thus, the higher-frequency light is not favorable for the photoinduced 120-degree spin ordered state. It is also noteworthy that for the higher electron filling of $n_{\rm e}=1/4$ in Fig.~\ref{Fig07}(c), the relative spin correlation $\Delta \hat{S}$ is suppressed when the light amplitude is too small as $E^\omega \ll 1$. This is because the ground-state magnetism in the system with $n_{\rm e}=1/4$ is not a ferromagnetic state but a four-sublattice noncollinear state, which is hardly destabilized by the light-induced deformation of the electronic structure and competes with the 120-degree spin ordered state tenaciously when $E^\omega$ is small.

\subsection{Appendix C. Light polarization}
\begin{figure*}[tb]
\centering
\includegraphics[scale=1.0]{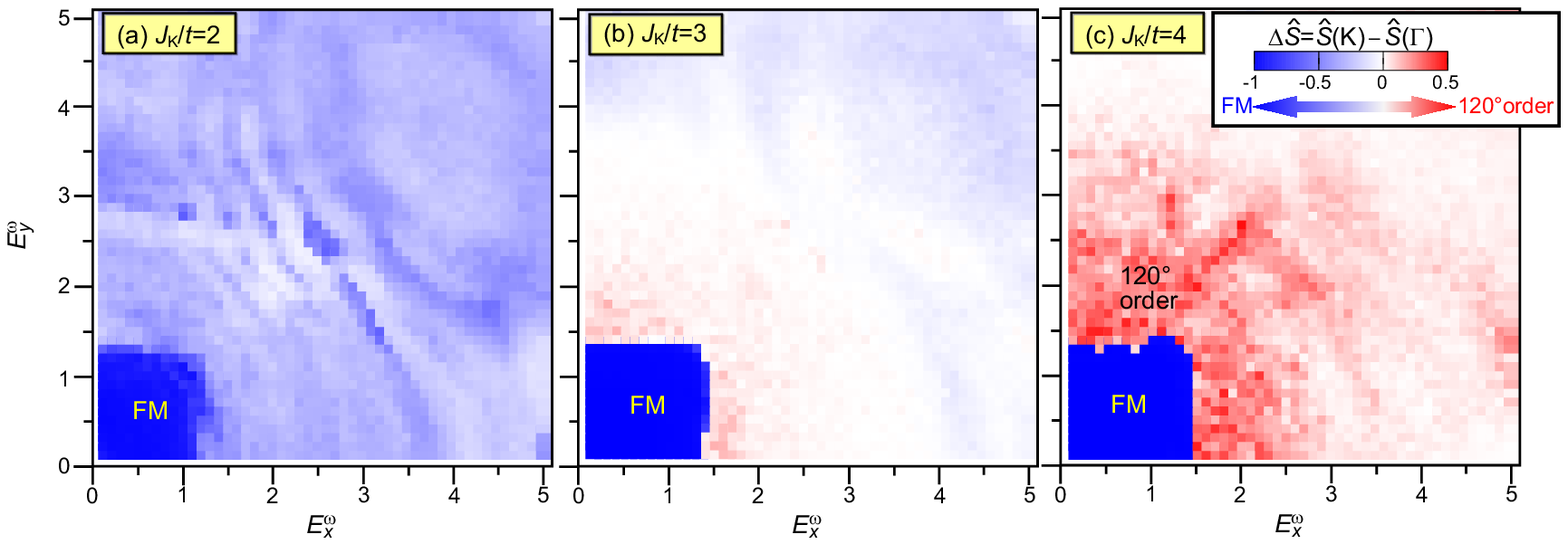}
\caption{Color maps of the relative spin correlation $\Delta \hat{S} \equiv \hat{S}({\rm K})-\hat{S}(\Gamma)$ for the photoirradiated Kondo-lattice model on a triangular lattice after sufficient duration, i.e., at $\tau=2000$, in plane of the $x$- and $y$-component amplitudes $E^\omega_x$ and $E^\omega_y$ of elliptically polarized light for various coupling strengths, i.e., (a) $J_{\rm K}/t=2$, (b) $J_{\rm K}/t=3$, and (c) $J_{\rm K}/t=4$. The calculations are performed for a system of $18 \times 18$ sites with periodic boundary conditions by setting $n_{\rm e}=1/9$, $\omega=1$ and $\alpha_{\rm G}=1$. The photoinduced 120-degree spin ordered state appears when the Kondo coupling is rather strong as $J_{\rm K}/t=3$ and $J_{\rm K}/t=4$ and the amplitude of light is larger than a certain threshold. Importantly, the photoinduced magnetic transition to the 120-degree spin order turns out to be induced even by linearly polarized light with $E_x^\omega=0$ or $E_y^\omega=0$.}
\label{Fig08}
\end{figure*}
We further investigate dependency on the light polarization by constructing color maps of the relative spin correlation $\Delta \hat{S}$ for the photodriven systems in plane of the $x$- and $y$-component amplitudes $E_x^\omega$ and $E_y^\omega$ of elliptically polarized light. Figure~\ref{Fig08} presents the calculated color maps for various strengths of $J_{\rm K}$, i.e., (a) $J_{\rm K}/t=2$, (b) $J_{\rm K}/t=3$, and (c) $J_{\rm K}/t=4$. The calculations are performed for the Kondo-lattice model with $n_{\rm e}=1/9$ on a triangular lattice of $18 \times 18$ sites with periodic boundary conditions by setting $\alpha_{\rm G}=1$ and $\omega=1$. Note that the light is of perfect circular polarization when $E_x^\omega=E_y^\omega$, whereas it is of linear polarization when $E_x^\omega=0$ or $E_y^\omega=0$. Noticeably, there are threshold amplitudes of the light electric field for activation of the magnetic system. When the light amplitude is as weak as $0 \le E_x^\omega \leq 1$ and $0 \le E_y^\omega \leq 1$, the nearly perfect ferromagnetic order with $\Delta \hat{S}\sim -1$ remains. The area and ranges of this inactive regime do not change so much upon the variation of Kondo-coupling strength $J_{\rm K}$.

On the contrary, behaviors beyond this inactive regime sensitively depend on the strength of Kondo coupling $J_{\rm K}$. When the Kondo coupling is rather weak as $J_{\rm K}/t=2$ [Fig.~\ref{Fig08}(a)], the relative spin correlation $\Delta \hat{S}$ is weakly negative even in the outer area of the inactive regime, indicating that the 120-degree spin ordered state never emerges. For a moderate strength of the Kondo coupling as $J_{\rm K}/t=3$ [Fig.~\ref{Fig08}(b)], the relative spin correlation $\Delta \hat{S}$ becomes weakly positive in a narrow area of $E^\omega \lesssim 2$ outside the inactive regime. Here $E^\omega=\sqrt{{E^\omega_x}^2+{E^\omega_y}^2}$ is the absolute amplitude of light. For a further increased strength of $J_{\rm K}/t=4$ [Fig.~\ref{Fig08}(c)], enhanced correlation of 120-degree spin order is observed as positive $\Delta \hat{S}$ in a wide area outside the inactive regime. Taking a look at the map in Fig.~\ref{Fig08}(c) more carefully, we realize that $\Delta \hat{S}$ takes large values when the absolute amplitude $E^\omega$ is less than $\sim 3$, while it is suppressed when $E^\omega$ exceeds this value, indicating that too intense light field is not suitable for the photoinduction of 120-degree spin order. This is again because the intense light field excites the conduction electrons in the lower band to the upper band above the band gap and thus works destructively for the pseudo half-filled electronic structure with uniformly occupied lower band and empty upper band required for the emergence of 120-degree spin order.

\subsection{Appendix D. Antiferromagnetic exchange coupling}
\begin{figure*}[tb]
\centering
\includegraphics[scale=1.0]{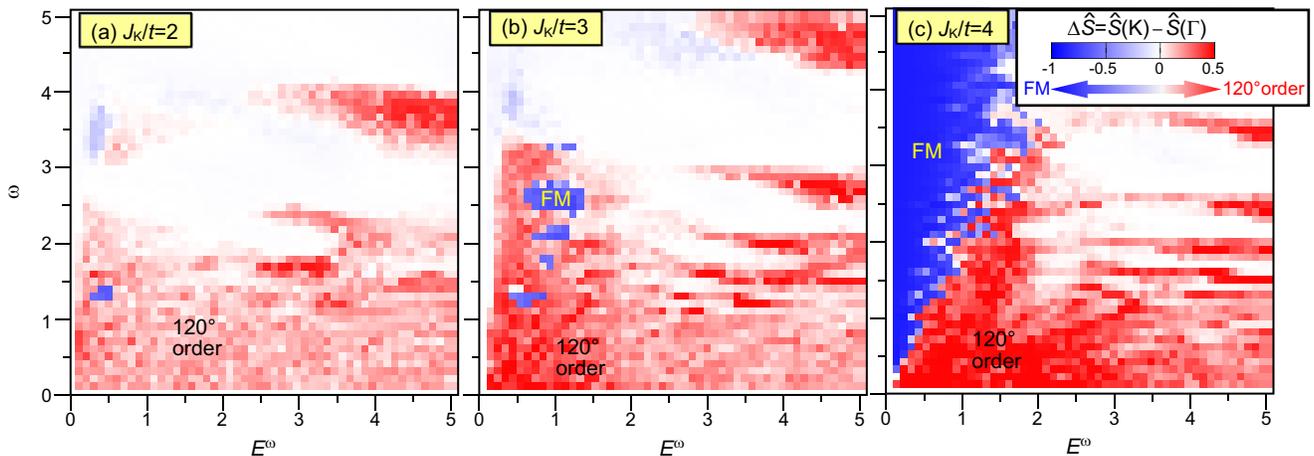}
\caption{Color maps of the relative spin correlation $\Delta \hat{S} \equiv \hat{S}({\rm K})-\hat{S}(\Gamma)$ for the photoirradiated Kondo-lattice model on a triangular lattice with tiny antiferromagnetic exchange coupling of $J_{\rm AFM}/t=0.01$ among adjacent localized spins after sufficient duration (at $\tau=2000$) in plane of the amplitude $E^\omega$ and the frequency $\omega$ of circularly polarized light for various Kondo-coupling strengths, i.e., (a) $J_{\rm K}/t=2$, (b) $J_{\rm K}/t=3$, and (c) $J_{\rm K}/t=4$. The calculations are performed for a system of $18 \times 18$ sites with periodic boundary conditions by setting $n_{\rm e}=1/6$ and $\alpha_{\rm G}=1$. The antiferromagnetic exchange coupling $J_{\rm AFM}$ favors the emergence of 120-degree spin ordered state. Comparing with the results in Fig.~\ref{Fig05} obtained without considering the antiferromagnetic exchange coupling, we find that the area of dominant 120-degree spin correlations significantly spreads over the $E^\omega$-$\omega$ plane, indicating that the coupling indeed supports the photoinduced phase transition to the 120-degree spin ordered state.}
\label{Fig09}
\end{figure*}
We also investigate the effect of antiferromagnetic exchange coupling among the localized spins $\{\bm S_i\}$. In the present study, we have examined the Kondo-lattice model without direct interactions among the localized spins up to now. However, it is natural to expect the presence of direct interactions in real materials. It is known that a ground-state magnetism is the 120-degree order for a classical Heisenberg model on the triangular lattice with the nearest-neighbor antiferromagnetic exchange coupling. Therefore, it is naively expected that incorporation of the antiferromagnetic exchange coupling should enhance the photoinduced 120-degree spin order. We examine its effect by adding the following term to the Hamiltonian,
\begin{align}
\mathcal{H}_\mathrm{AFM} = J_\mathrm{AFM}\sum_{\langle ij \rangle}\bm{S}_i\cdot\bm{S}_j,
\label{Eq24}
\end{align}
with a very tiny coupling strength of $J_{\rm AFM}/t=0.01$.

Figure~\ref{Fig09} presents calculated color maps of the relative spin correlation $\Delta \hat{S}$  for the photodriven systems in plane of the amplitude $E^\omega$ and the frequency $\omega$ of circularly polarized light for various strengths of $J_{\rm K}$, i.e., (a) $J_{\rm K}/t=2$, (b) $J_{\rm K}/t=3$, and (c) $J_{\rm K}/t=4$. The calculations are performed for the Kondo-lattice model with $n_{\rm e}=1/6$ on a triangular lattice of $18 \times 18$ sites with periodic boundary conditions by setting $\alpha_{\rm G}=1$. Note that the ground state in the equilibrium system is ferromagnetic even with this tiny antiferromagnetic coupling. Comparison of the spin-correlation maps when $J_{\rm AFM}=0$ (Fig.~\ref{Fig06}) and those when $J_{\rm AFM}/t=0.01$ (Fig.~\ref{Fig09}) reveals that the tiny antiferromagnetic exchange coupling significantly enhances the stability of nonequilibrium 120-degree spin order under photoirradiation. The region of dominant 120-degree spin correlation spreads over a wide area in the $E^\omega$-$\omega$ plane. Note that the smallest values of $E^\omega$ and $\omega$ examined in the present simulations are 0.1 (2 MV/cm) and 0.1 (24.2 THz), respectively. The spin-correlation maps in Fig.~\ref{Fig09} indicate that a light field of $E^\omega=2$ MV/cm or even weaker light fields can induce the 120-degree spin order as far as the weak antiferromagnetic exchange coupling exists. This finding supports the experimental feasibility of the proposed photoinduced magnetic phase transition as will be discussed later.

Another important aspect seen in Fig.~\ref{Fig09} is that the effect of the antiferromagnetic coupling is pronounced when the Kondo coupling is weak. Namely, when the Kondo coupling is as weak as $J_{\rm K}/t=2$, the ferromagnetic correlation is strongly suppressed, while the 120-degree spin order is significantly stabilized as seen in the comparison with Fig.~\ref{Fig06}(a) and Fig.~\ref{Fig09}(a). On the contrary, the boundary and areas of these two phases do not change so much when the Kondo coupling is strong as $J_{\rm K}/t=4$ as seen in the comparison with Fig.~\ref{Fig06}(c) and Fig.~\ref{Fig09}(c). This finding is surprising because in the absence of the antiferromagnetic coupling, a stronger Kondo coupling is favorable for the emergence of 120-degree spin order under photoirradiation as argued above (see also Fig.~\ref{Fig06} and Fig.~\ref{Fig08}). This aspect indicates that the mechanisms for stabilizing the photoinduced 120-degree spin order are different between the Kondo coupling $J_{\rm K}$ and the antiferromagnetic coupling $J_{\rm AFM}$, and they do not work cooperatively.

\subsection{Appendix E. Experimental feasibility}
\begin{figure*}[tb]
\centering
\includegraphics[scale=1.0]{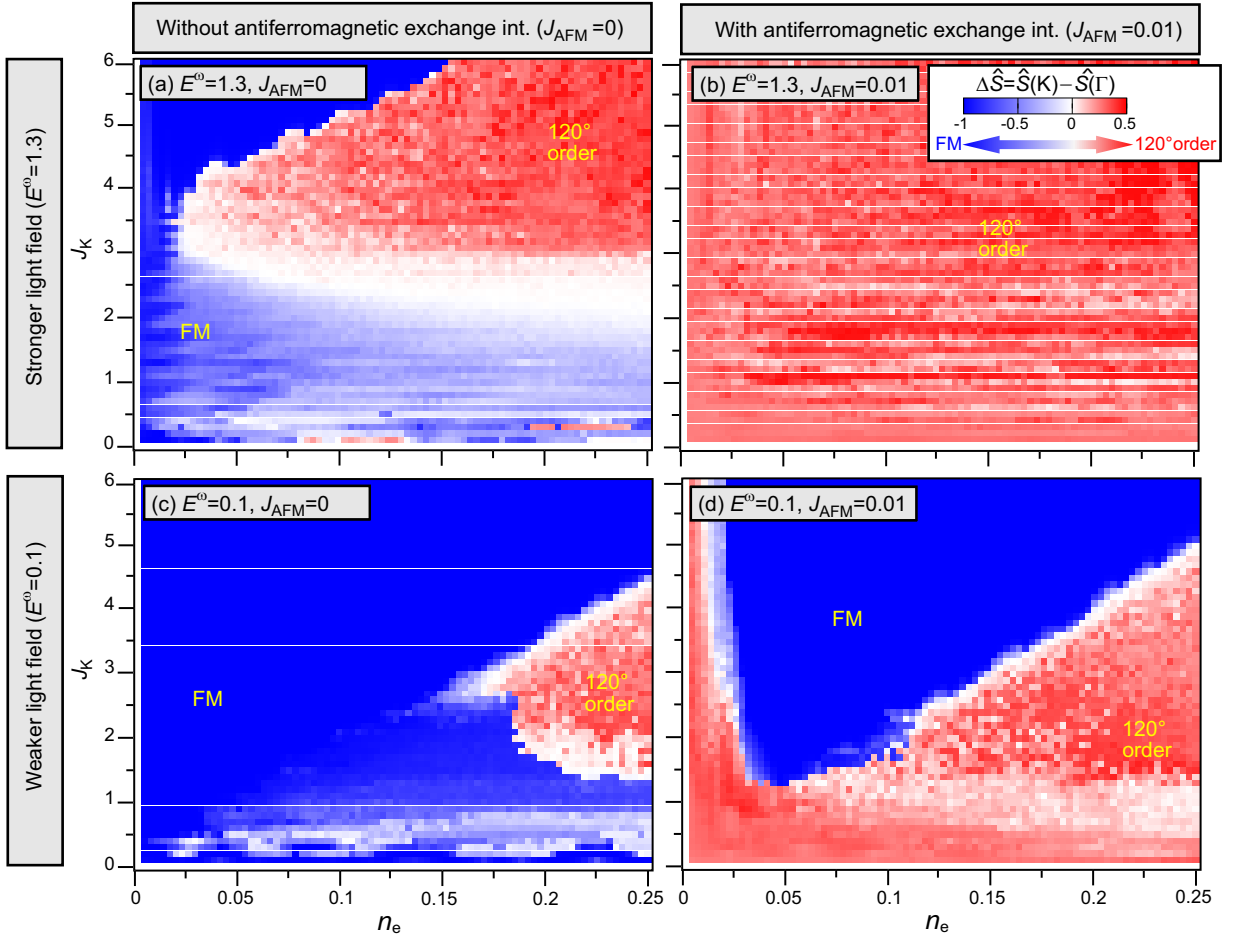}
\caption{Color maps of the relative spin correlation $\Delta \hat{S} \equiv \hat{S}({\rm K})-\hat{S}(\Gamma)$ for the photoirradiated Kondo-lattice model on a triangular lattice after sufficient duration (at $\tau=2000$) in plane of the electron filling $n_{\rm e}$ and the strength of Kondo coupling $J_{\rm K}$. The calculations are performed for a system of $18 \times 18$ sites with periodic boundary conditions by setting $\alpha_{\rm G}=1$. (a) Color map of $\Delta \hat{S}$ in the $n_{\rm e}$-$J_{\rm K}$ plane obtained for stronger light field of $E^\omega=1.3$ and $\omega=0.8$ in the absence of antiferromagnetic exchange coupling ($J_{\rm AFM}=0$). (b) Color map obtained for stronger light field of $E^\omega=1.3$ and $\omega=0.8$ in the presence of tiny antiferromagnetic exchange coupling of $J_{\rm AFM}/t=0.01$. (c) Color map of $\Delta \hat{S}$ for weak light field of $E^\omega=0.1$ and $\omega=0.8$ when $J_{\rm AFM}=0$. (d) Color map of $\Delta \hat{S}$ for weak light field of $E^\omega=0.1$ and $\omega=0.8$ when $J_{\rm AFM}/t=0.01$.}
\label{Fig10}
\end{figure*}
We finally discuss experimental feasibility of the predicted photoinduced phenomenon. For this purpose, we construct color maps of the relative spin correlation $\Delta \hat{S}$ for the photodriven systems in plane of the electron filling $n_{\rm e}$ and the strength of Kondo coupling $J_{\rm K}$ for four different combinations of the cases, i.e., the cases of strong and weak light electromagnetic fields $E^\omega$ and the cases with and without antiferromagnetic exchange coupling $J_{\rm AFM}$. Concerning the intensity of light electromagnetic field, we consider $E^\omega=1.3$ for the strong light field and $E^\omega=0.1$ for the weak light field, which correspond to 26 MV/cm and 2 MV/cm in real units, respectively, when we assume a transfer integral $t=1$ eV and a lattice constant $a=5$ \AA. The former intensity is so strong that it is rather hard to realize in experiments, whereas the latter intensity is relatively feasible. Concerning the antiferromagnetic exchange coupling, we assume a very weak coupling strength of $J_{\rm AFM}/t=0.01$ when it is present, while we set $J_{\rm AFM}=0$ when it is absent. Note that in this paper, we have mainly discussed results obtained for the strong light field of $E^\omega=1.3$ to see the physical phenomenon clearly. In the following, we will discuss that the phenomenon can be realized even with easily accessible light intensities. All the calculations are performed for a system of $18 \times 18$ sites with periodic boundary conditions by setting $\alpha_{\rm G}=1$. We adopt circularly polarized light with a time-dependent vector potential $\bm A(\tau)$ given by Eq.~(\ref{eq:Eq23}).

We first find in Fig.~\ref{Fig10}(a) that the photoinduced 120-degree spin order emerges in a wide area in the $n_{\rm e}$-$J_{\rm K}$ plane when the light field is strong as $E^\omega=1.3$, which spreads even to a low-filling region of $n_{\rm e}\sim0.03$. When we introduce a tiny antiferromagnetic coupling of $J_{\rm AFM}/t=0.01$, the region further spreads over entire area of the plane as seen in Fig.~\ref{Fig10}(b). Note that the smallest values of $J_{\rm K}$ and $n_{\rm e}$ examined in the present simulations are 0.1 and $1/N$ ($N=18^2$ is the system size), respectively. These results indicate that the photoinduced 120-degree spin order emerges even when the Kondo coupling is very weak and the electron filling $n_{\rm e}$ is very small. Then what happens when the light field becomes one order magnitude weaker? Surprisingly, we find in Fig.~\ref{Fig10}(c) that the 120-degree spin order appears still in reasonable parameter ranges of $n_{\rm e}$ and $J_{\rm K}$. When the tiny antiferromagnetic coupling $J_{\rm AFM}$ is introduced, the area of the dominant 120-degree spin correlation becomes larger as seen in Fig.~\ref{Fig10}(d). In real magnets, the localized spins are usually interacting with each other at least weakly, and it is natural to expect the presence of antiferromagnetic coupling as weak as $J_{\rm AFM}/t=0.01$. Importantly, this weak antiferromagnetic coupling can drastically change the situation. Namely, a weak light field of $E^\omega=0.1$ can induce the 120-degree spin order even when the Kondo coupling $J_{\rm K}$ is considerably weak and the electron filling $n_{\rm e}$ is away from the half filling. These aspects strongly support the experimental feasibility to observe the predicted photoinduced 120-degree spin order in spin-charge coupled magnets with triangular crystal structure.

One candidate material to observe the predicted photoinduced spin order is CeRh$_6$Ge$_4$~\cite{Matsuoka2015,BShen2020,WuY2021,Kotegawa2019}, which exhibits a ferromagnetic ground state on a triangular lattice and is expected to be described by the ferromagnetic Kondo-lattice model. Another interesting material is Gd$_2$PdSi$_3$~\cite{Kurumaji2019}, which is also expected to be described by the Kondo-lattice model on the triangular lattice. Although the ground-state magnetism is not ferromagnetic but spiral, we expect a photoinduced magnetic transition in this material due to the similar mechanism, i.e., the optical modulation of electronic structure.

\subsection{Appendix F: Gilbert damping}
\begin{figure}[tb]
\centering
\includegraphics[scale=1.0]{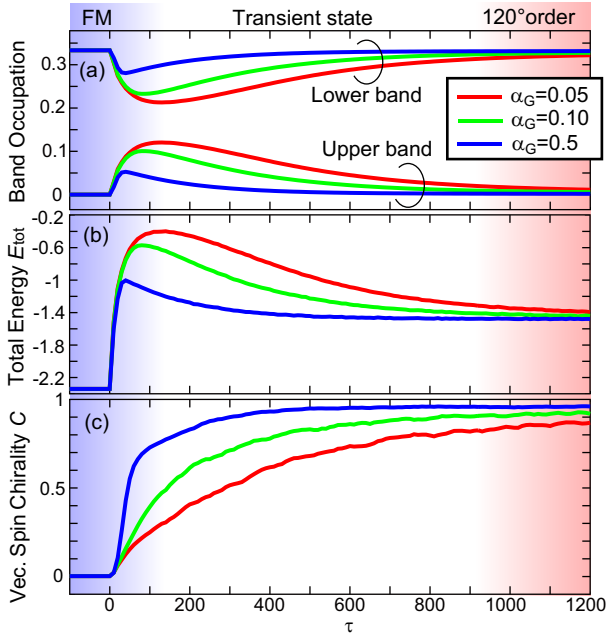}
\caption{Calculated time profiles of (a) number of electrons in the lower band and that in the upper band (divided by the total number of sites $N$), (b) total energy $E_{\rm tot}$, and (c) averaged vector spin chirality $C$ for various values of the Gilbert-damping coefficient, i.e., $\alpha_{\rm G}$=0.05, 0.10 and 0.5. The calculations are performed for the Kondo-lattice model with $J_{\rm K}/t=4$ and $n_{\rm e}=1/6$ on a triangular lattice with $18\times18$ sites irradiated by circularly polarized light of $E^\omega=1.3$ and $\omega=0.8$.}
\label{Fig11}
\end{figure}
In the present paper, we have mainly studied the cases with $\alpha_{\rm G}=0.1$ and $\alpha_{\rm G}=1$. Here we discuss what is expected to happen when $\alpha_{\rm G}$ varies. We have examined the cases with various values of $\alpha_{\rm G}$ and have found that the effect is simple, that is, the duration of photoirradiation required for the emergence of 120-degree spin order increases as $\alpha_{\rm G}$ decreases. When we start irradiating the system, the conduction electrons are first excited to the upper band from the lower band. The excited electrons subsequently fall to the lower band through the relaxation process to realize the 120-degree spin order because of the energy dissipation due to the Gilbert damping. Importantly, it takes longer to reach this 120-degree spin order as a nonequilibrium steady phase when the dissipation is weaker with a smaller $\alpha_{\rm G}$. Calculated time profiles of several quantities in Fig.~\ref{Fig11} clearly demonstrate this aspect. Namely, temporal evolutions of the electron occupations of upper and lower bands, the total energy, and the averaged vector spin chirality converges more slowly for a smaller $\alpha_{\rm G}$.

\section{Acknowledgement}
We thank Atsushi Ono for useful discussions and Yasuhiro Tanaka for his instructive help in the numerical simulations. This work is supported by Japan Society for the Promotion of Science KAKENHI (Grant No. 16H06345, No. 19H00864, No. 19K21858, and No. 20H00337), CREST, the Japan Science and Technology Agency (Grant No. JPMJCR20T1), a Research Grant in the Natural Sciences from the Mitsubishi Foundation, and a Waseda University Grant for Special Research Projects (Project No. 2020C-269 and No. 2021C-566). 

\end{document}